# Resolution of the St. Petersburg paradox using Von Mises' axiom of randomness

Andrea Berdondini

ABSTRACT: In this article we will propose a completely new point of view for solving one of the most important paradoxes concerning game theory. The method used derives from the study of non-ergodic systems. This circumstance may create a dependency between results that are often extremely difficult to detect and quantify, such as in the field of finance. Consequently, the expected gain obtained from data that may be correlated has a statistical value that is difficult to determine, thus it cannot be used for decision-making purposes. Therefore, in this scenario, an alternative parameter to be use during the decision-making process must be found. The solution develop shifts the focus from the result to the strategy's ability to operate in a cognitive way by exploiting useful information about the system. In order to determine from a mathematical point of view if a strategy is cognitive, we use Von Mises' axiom of randomness. Based on this axiom, the knowledge of useful information consequently generates results that cannot be reproduced randomly. Useful information in this case may be seen as a significant datum for the recipient, for their present or future decision-making process. In conclusion, the infinite behaviour in this paradox may be seen as an element capable of rendering the expected gain unusable for decision-making purposes. As a result, we are forced to face the problem by employing a different point of view. In order to do this we shift the focus from the result to the strategy's ability to operate in a cognitive way by exploiting useful information about the system. Finally, by resolving the paradox from this new point of view, we will demonstrate that an expected gain that tends toward infinity is not always a consequence of a cognitive and non-random strategy. Therefore, this result leads us to define a hierarchy of values in decision-making, where the cognitive aspect, whose statistical consequence is a divergence from random behaviour, turns out to be more important than the expected gain.

## Introduction

The St. Petersburg paradox represents one of the most important paradoxes in game theory. The classic solution used to solve uses special utility functions that implement the concept of marginal utility [1], [2], [3]. This type of approach has been strongly criticized in virtue of the fact that utility functions attempt to formalize sociological behaviour from a mathematical point of view, and this is why they always have a subjectivity component. Moreover, many studies of behavioural economics [4], [5] highlight how people's behaviour is often irrational. Consequently, the resolution of this paradox still represents an open challenge and, as we will see, the search for an alternative solution may help us improve the decision-making process during the evaluation of a strategy.

In order to understand the method proposed in this article for resolving the St. Petersburg paradox, we must first explain the origins of this method. This approach was developed to study the strategies operating on non-ergodic systems. In particular, the primary field of application is characterized by the study of quantitative trading algorithms operating on financial markets. The non-ergodicity condition can make the results dependent on each other. Therefore, in this scenario a dependency is created between data which, like in the field of finance, is difficult to detect and quantify.

To explain to you the possible consequences at a decision-making level of this condition, I propose the following example: think about making a hundred bets on a hundred flips of a coin and winning one hundred times. In this case, you will have obtained a hundred victories independent of each other to which a very high-expected gain will be associated. You therefore reach the right conclusion that the strategy used to predict the flip of the coin is most likely correct. Now let's change the starting scenario and let's say we make 100 equal bets on a single coin flip, getting a hundred wins. Obviously, since the bets are completely dependent on each other, in this case they cannot be used to calculate the expected gain. In fact, it's basically as if we made a single bet. Now let's imagine, as a third and last scenario, that you are not able to see the person flipping the coin: if we win all the hundred times, we don't know if they are dependent or independent of each other. This third scenario generates a very important decision-making problem, because if I consider the results as independent and they are not, I risk overestimating the strategy. This wrong assessment can lead me to make the irrational choice of using a useless strategy. In finance the non-independence of the results creates a statistical phenomenon called clustering. This statistical characteristic determines the formation of groups of high returns, alternating with groups of low returns. In other words, it means that returns are not distributed evenly but tend to cluster together. The clustering phenomenon has disastrous effects in finance, because when you are going through winning phases you are led to consider the operations carried out as independent of each other. This implies that the expected gain, calculated from data that we mistakenly think to be independent, is overestimated, therefore the evaluation of the strategy will also be incorrect. So, this behaviour can subject us to unexpected risk. With regards to this topic we have developed a paradox [6], which we have called "the professional trader's paradox". The name derives from the fact that we are inclined to consider that our operations are always independent, and therefore, when we face a series of winning bets, we tend to overestimate the strategy used.

In conclusion, the expected gain, obtained from data that may not be independent, cannot be used for decision-making purposes, as it has a statistical value that is difficult to determine. Therefore, from this example we understand that there are situations, like in non-ergodic systems, where the expected gain is no longer a reliable parameter. Consequently, we can think that other situations may exist, like in the case of the infinite behaviour of this paradox, where the expected gain is a datum that cannot be used in decision-making.

We begin to understand that the problem in resolving the St. Petersburg paradox may derive from considering the expected gain, and its variants (utility functions), as the only possible point of view in the evaluation of a strategy. So, the question we have to ask ourselves is: is there a parameter that is better than the expected gain?

The answer we give to this question is focused on being able to understand, from a statistical point of view, if a strategy operates in a cognitive way by exploiting useful information present on the system. The useful information in this case can be seen as a datum subject to analysis that rendered it significant to the recipient for their present or future decision-making process. To determine mathematically if a strategy is cognitive, in the sense just described, we exploit the Von Mises' axiom of randomness. The axiom defines the statistical characteristic that must have a sequence in order to be considered random. The axiom is the following: "the essential requirement for a sequence to be defined as random consists in the complete absence of any rules that may be successfully applied to improve predictions about the next number ".

The meaning of this axiom is the following: when we understand a set of rules to which a numerical sequence is subject we can obtain results, intended as forecasts on the next number of the sequence, whose probability of being reproduced randomly tends toward zero on increasing the number of forecasts made.

Consequently, the results obtained with a game strategy that implements information useful to improve our probability of winning, generates results that cannot be reproduced randomly. Basically, the probability of obtaining better results with a random strategy compared to a cognitive strategy, which implements useful information, tends toward zero as the number of predictions made increases.

This axiom is indeed a statistical method for evaluating the results obtained, without taking into consideration the absolute value of the expected gain. In fact, this method is based solely on the fact of being able to discriminate whether the results were obtained with a random strategy or through a cognitive strategy that implements a set of rules to which the system is subject.

In this way we obtain a fundamental result in analysing the strategies of a particular class of zero-sum games, where there is a balance between the participants. Balance means the situation where none of the players has an implicit advantage over the others. The famous mathematician Daniel Bernoulli defined this particular class of games as: "mathematically fair game of chance".

This type of game plays a particularly important role in game theory, because it represents a very frequent situation in various fields of interest such as finance.

If we analyse the results obtained by repeating the game of chance described in the St. Petersburg paradox a large number of times, in the next paragraph we will demonstrate that better profits can be obtained with a probability that tends toward 50% by using a random strategy. In practice, the results of a purely random game strategy tend to be distributed symmetrically with respect to the expected theoretical gain derived from the strategy described in the paradox. This result indicates that the doubling-down strategy after each lost bet does not exploit any kind of useful information, and therefore it is a completely non-cognitive game method. Consequently, by taking the cognitive aspect as a parameter to be used in decision-making, and having demonstrated the complete absence within the strategy, we are able to solve the paradox by proving the irrationality of the game method.

In this article, we want to introduce the cognitive aspect, understood in the sense of acting in a non-random way by exploiting useful information about the system, as a fundamental element for improving the decisions theory. In fact, this paradox is useful to make us understand that the knowledge of useful information about the system, capable of increasing our probability of victory, always involves an increase of the expected gain. However, the opposite is not true: an expected gain that tends to infinity does not imply that the strategy exploits knowledge about the system and therefore is cognitive and not random. Consequently, a hierarchy of values is created, where the cognitive aspect is more important than the expected gain for decision-making purposes.

**Resolution of the St. Petersburg paradox**

In this paragraph, we will solve the St. Petersburg paradox by demonstrating that the doubling-down strategy after each lost bet is a non-cognitive strategy, which implements no useful information that can be used to improve the probability of success.

In order to do this we have to define the random strategy, which we will use to calculate the probability of obtaining better results than those obtained with the gambling method defined in the paradox. In fact, as mentioned in the previous paragraph, this probability should tend to zero if the strategy being evaluated is a cognitive strategy that implements useful information. Firstly, we define some parameters that are fundamental to characterize our random strategy of reference.

The first parameter we need is the expected value EV of the game obtained by using the doubling-down strategy after each bet lost. Given a number of flips equal to L, with a 50% probability of winning and placing the value of the first bet equal to 1, we have:

$$EV = \frac{L}{2}$$

The second parameter is the average bet AB. By carrying out L bets of Bn value, we have:

$$AB = \frac{(B1 + B2 \ldots + BL)}{L}$$

Knowing that the first bet B1 is equal to 1, and double downing after every bet lost and returning to value 1 when we win the bet, we have:

$$AB = \sum_{n=2}^{L} \frac{(L-n+1)2^{n-1}}{2^n L} + 1$$

$$AB = \sum_{n=2}^{L} \frac{(L-n+1)}{2L} + 1$$

$$AB = \frac{L(L-1)}{2} \frac{1}{2L} + 1$$

$$AB = \frac{L-1}{4} + 1$$

At this point the random game strategy will be defined as follows: given a number of flips equal to L, L bets of AB constant value will be made, randomly choosing whether to bet heads or tails on each bet. To calculate whether a strategy of this type can obtain better results compared to the expected value EV of the strategy of the paradox, just use the binomial distribution formula.

$$F(k, L, p) = \binom{L}{k} p^k (1-p)^{L-k}$$

P = probability of winning

K = number of wins

L = number of tosses

By using the binomial distribution formula, given a value of L, we can obtain the probability of achieving better results with the random strategy described above. The results for the L values ranging from 10 to 200 are shown in Figure 1. Looking at the figure, we see how the probability tends asymptotically toward 50%. Therefore, we have a 50% chance of getting better or worse results. Basically, the strategy described in the paradox tends asymptotically toward a random strategy. Consequently, the doubling-down strategy turns out to be a strategy that does not implement useful information for improving our likelihood of victory. Thus, using the cognitive aspect as a method of evaluation has proven the irrationality of the strategy.

We use very similar approaches, where the strategy being evaluated is compared with an equivalent random strategy, in the financial field to analyse the results generated by a trading strategy [7], [8].

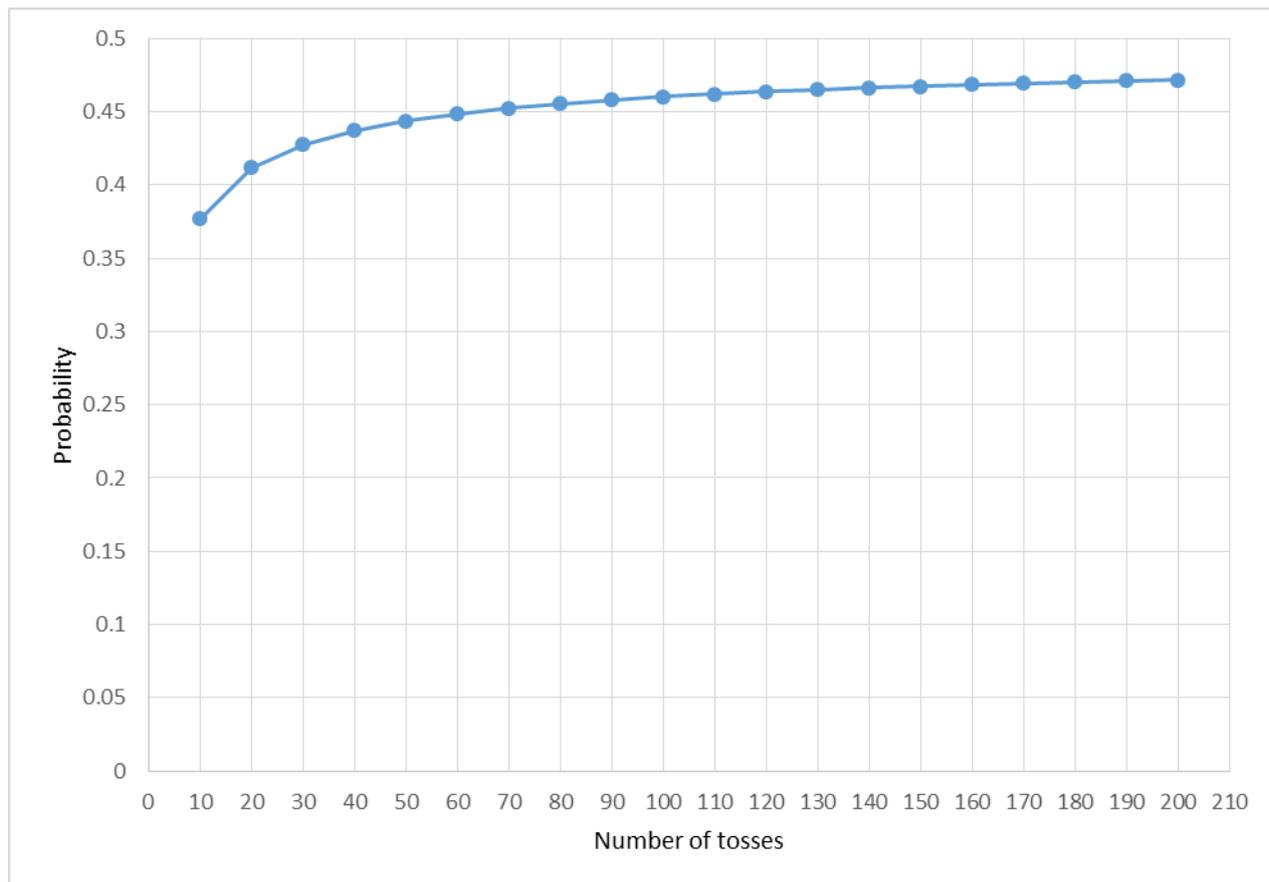

**FIG. 1**: Probability of obtaining better results with a random strategy, given a number of L tosses ranging from 10 to 200.

## Conclusion

In this article we use the St. Petersburg paradox to introduce a parameter related to the cognitive aspect of a strategy, as a fundamental element to help our decision-making in all those situations where the expected gain turns out to be an unreliable parameter. This approach was developed by studying non-ergodic systems. In this scenario the results can be non-independent, so the expected gain becomes a parameter with a statistical value that is difficult to determine. Therefore, it cannot be used in decision-making and a new parameter needs to be found. The parameter chosen is related to the strategy's ability to operate in a cognitive way (the cognitive term indicates the strategy's ability to operate in a non-random way by exploiting useful information about the system, capable of making us increase the probability of victory).

To determine mathematically if a strategy is cognitive, we used the von Mises' axiom of randomness. Based on this axiom, strategies that implement useful information about the system generate results that cannot be reproduced randomly. Thus, we compared the paradox strategy with a completely random but equivalent strategy from the point of view of the total betting value. From this comparison, we have demonstrated that the random strategy gets better results with a probability that tends toward 50% as the number of tosses increases. Basically, the strategy tends to converge to a random strategy instead of diverging as we would expect from a cognitive strategy. In fact, if a strategy implements useful information on the system, the probability of randomly obtaining better results tends toward zero. This result indicates that the doubling-down strategy after each lost bet is not a cognitive strategy that exploits useful information about the system, and therefore by taking the cognitive aspect as an evaluation parameter we have solved the paradox.

In conclusion, the St. Petersburg paradox teaches us that an expected gain that tends toward infinity does not always imply the presence of a cognitive and non-random strategy. Thus knowledge, meaning the exploitation of useful information capable of making us increase the probability of victory, always implies an increase in the expected gain, but the opposite is not true; an expected gain that tends toward infinity can also be obtained in the absence of knowledge about the system. Consequently, from the decision-making aspect we can create a hierarchy of values, where knowledge is more important than the expected gain. In fact, the expectation of victory can be difficult to estimate as in the case of non-ergodic systems or be a non-useful data if the developed strategy has a high degree of overfitting. In all these cases the calculation of the probability of obtaining the same results randomly becomes a much more reliable parameter, since this datum is influenced only by the real knowledge we have of the system and not by the noise. In fact, a statistical datum does not represent a useful information, but becomes a useful information only when it is possible to proof that it was not obtained in a random way. In practice, the probability of obtaining the same result randomly must be very low in order to consider the result useful.

*E-mail address*: andrea.berdondini@libero.it